\documentclass[aps,prl,longbibliography,twocolumn,superscriptaddress,amsfont,graphicx,nofootinbib,preprintnumbers]{revtex4-1}%
\UseRawInputEncoding
\usepackage{color,graphicx,epsfig}
\usepackage{ifpdf}
\usepackage{amsmath}
\usepackage{bm}
\usepackage[english]{babel}
\usepackage{amssymb}
\usepackage{braket}
\usepackage{setspace}
\usepackage{multirow}
\allowdisplaybreaks[4]

\usepackage{hyperref}
\usepackage{enumerate}
\usepackage{lipsum}

\usepackage{slashed}

\usepackage{changes}

\begin{document}
\title{Direct Detection of Leptophobic Dark Matter with Electronic Collective Excitations}


\author{Yuanlin Gong}
\email{yuanlingong@nnu.edu.cn}
\affiliation{Department of Physics and Institute of Theoretical Physics, Nanjing Normal University, Nanjing, 210023, China}

\author{Yifan Hu}
\email{yifanhu@nnu.edu.cn}
\affiliation{Department of Physics and Institute of Theoretical Physics, Nanjing Normal University, Nanjing, 210023, China}

\author{Ning Liu}
\email{liuning@njnu.edu.cn}
\affiliation{Department of Physics and Institute of Theoretical Physics, Nanjing Normal University, Nanjing, 210023, China}
\affiliation{Nanjing Key Laboratory of Particle Physics and Astrophysics, Nanjing, 210023, China}

\author{Liangliang Su}
\email{liangliang.su@kit.edu}
\affiliation{Institute for Astroparticle Physics (IAP), Karlsruhe Institute of Technology (KIT), Hermann-von-Helmholtz-Platz 1, D-76344 Eggenstein-Leopoldshafen, Germany}

\author{Bin Zhu}
\email{zhubin@mail.nankai.edu.cn}
\affiliation{Department of Physics, Yantai University, Yantai 264005, China}

\date{\today}

\begin{abstract}
Some new-generation dark matter detection experiments are primarily designed to search for the dark matter-electron interactions, but they can also be utilized to probe models in which dark matter couples exclusively to nucleon via the quantum effects. The hadronic loop-induced interactions can directly excite plasmons in semiconductors, thereby providing an additional channel for detecting the leptophobic dark matter. In this work, we investigate plasmon excitations in silicon detectors induced by boosted dark matter and cosmic-ray up-scattering dark matter via the hadronic loop process. By analyzing the available experimental data, we derive new exclusion limits on the leptophobic dark matter-nucleon scattering cross section in the sub-MeV mass range.   

\end{abstract}
\maketitle
\section{ Introduction }

A wealth of gravitational evidence from astrophysical and cosmological observations compellingly supports the existence of dark matter (DM), which constitutes the majority of our Universe’s mass. Therefore, the various detection experiments and DM candidates have been proposed to better understand DM. However, the fundamental nature and composition of DM particularly its non-gravitational interaction and mass is still unknown. For instance, the most widely studied DM candidate, the Weakly Interaction Masssive Particles (WIMPs)~\cite{Lee:1977ua} with mass from $\mathcal{O}(1)$ GeV to $\mathcal{O}(10)$ TeV, has not yielded any conclusive signals in direct detection experiments~\cite{PandaX:2024qfu,LZ:2024zvo,XENON:2025vwd}. Meanwhile, the DM community have increasingly focused on non-WIMP DM candidates, such as sub-GeV DM~\cite{Essig:2011nj,Hochberg:2015pha,Hochberg:2016ajh,Derenzo:2016fse,Ibe:2017yqa,Dolan:2017xbu,Knapen:2017xzo,Essig:2019xkx,Flambaum:2020xxo,Elor:2021swj,Su:2021jvk,Calabrese:2022rfa,Liang:2022xbu,Li:2022acp,Gu:2022vgb,Su:2022wpj,PandaX:2022xqx,Liang:2024lkk,Balan:2024cmq,Bhattiprolu:2024dmh,Lin:2025pqh,Griffin:2025eqm,Cheek:2025nul,Li:2025zwg}, and on the development of novel detector materials~\cite{Graham:2012su,Essig:2015cda,Hochberg:2016sqx,Knapen:2017ekk,Kurinsky:2019pgb,Trickle:2019ovy,Coskuner:2019odd,Geilhufe:2019ndy,Griffin:2020lgd,Andersson:2020uwc,Esposito:2022bnu,QUEST-DMC:2023nug,DELight:2024bgv,QROCODILE:2024zmg}, including semiconductor-based detectors~\cite{DAMIC:2016lrs,SENSEI:2020dpa,EDELWEISS:2020fxc,SuperCDMS:2020ymb,Oscura:2022vmi, CDEX:2022kcd,SENSEI:2024yyt,DAMIC-M:2025luv,Abbamonte:2025guf}.

The traditional DM direct detection experiments typically target nuclear recoil signals, whereas the semiconductor experiments focus on electronic excitation signals induced by the DM-electron interaction, which can search for the lighter DM. For the semiconductor, the DM-electron scattering rate can be described by the energy loss function (ELF) of the material, $\mathrm{Im}[-\epsilon^{-1}(\mathbf{Q},\omega)]$, which depends on the the momentum and frequency dependent longitudinal dielectric function $\epsilon(\mathbf{Q},\omega)$. It can be calculated by the density functional theory including the many-body effect in Ref.~\cite{Knapen:2021run,Knapen:2021bwg,Boyd:2022tcn,Dreyer:2023ovn}. The ELF displays a significant enhancement for the excitation rate in some certain kinematics region, such as $|\mathbf{Q}| < 5$ keV and $\omega \sim 15$ eV for the silicon detector as shown in Ref.~\cite{Knapen:2021bwg}. This phenomenon is so-called the electronic collective or plasmon excitation, which is a common feature for the well-ordered solid materials. The plasmon excitation region of some conventional and unconventional material can be found in the Ref.~\cite{Kurinsky:2020dpb} and \cite{Hochberg:2025dom}, respectively. Recently, Refs.\cite{Liang:2024xcx,Essig:2024ebk,Sun:2025gyj,Dent:2025drd} investigated the potential of directly plasmon excitations as a probe of DM-electron scattering. In particular, Ref.\cite{Liang:2024xcx} clearly demonstrated that direct plasmon excitation requires the kinematic condition $v_{\chi} \gtrsim 0.01c$, and further calculated the corresponding excitation rate induced by relativistic DM-electron interactions.

Previous studies of direct electronic collective excitations have primarily focused on DM-electron interactions, such as those mediated by a dark photon or within leptophilic DM frameworks. In contrast, for well-motivated scenarios in which DM couples only to nucleons, direct electronic signals cannot be generated at tree level. Nevertheless, such DM can directly produce observable electron signals through hadronic loop-induced interactions with electrons, thereby enabling the detection of lighter DM, as demonstrated in Ref.\cite{Diamond:2023fsm}. Motivated by this, in this work we investigate direct electronic collective excitations induced by leptophobic DM (LDM)\cite{Batell:2014yra,Coloma:2015pih,Dror:2017ehi,Berlin:2018bsc,Boyarsky:2021moj} via a proton-loop process.

Similarly, in this work, we first identify the kinematic requirements for direct plasmon excitation by LDM, which similarly demand the speed of LDM larger than $0.01c$. As benchmark scenarios, we consider two relativistic LDM fluxes, two-component boosted DM (BDM)~\cite{Agashe:2014yua,Fornal:2020npv,Borah:2021jzu,Basu:2023wgo,Li:2023fzv,Choi:2024ism, Alhazmi:2025nvt} and cosmic-ray up-scattering DM (CRDM)~\cite{Bringmann:2018cvk,Ema:2018bih,Cappiello:2019qsw,Wang:2019jtk,Ge:2020yuf,Xia:2020apm,Ema:2020ulo,Bell:2021xff,Wang:2021nbf,PandaX-II:2021kai,Maity:2022exk,Nagao:2022azp,Bell:2023sdq,Su:2023zgr,PandaX:2024pme,Herbermann:2024kcy,Diurba:2025lky,Gustafson:2025dff,Chen:2025twc}. We then recalculate the plasmon excitation rate from LDM-electron scattering mediated via hadronic loops, and use data from the SENSEI experiment at SNOLAB~\cite{SENSEI:2023zdf} to estimate the corresponding detection sensitivity in the sub-MeV mass range. This analysis yields new exclusion limits on the spin-independent LDM-nucleon scattering cross section through a light vector mediator.

\section{Leptophobic Dark Matter-electron scattering in semiconductor}

\begin{figure}[ht]
    \centering
    \includegraphics[height=6.5cm,width=6.5cm]{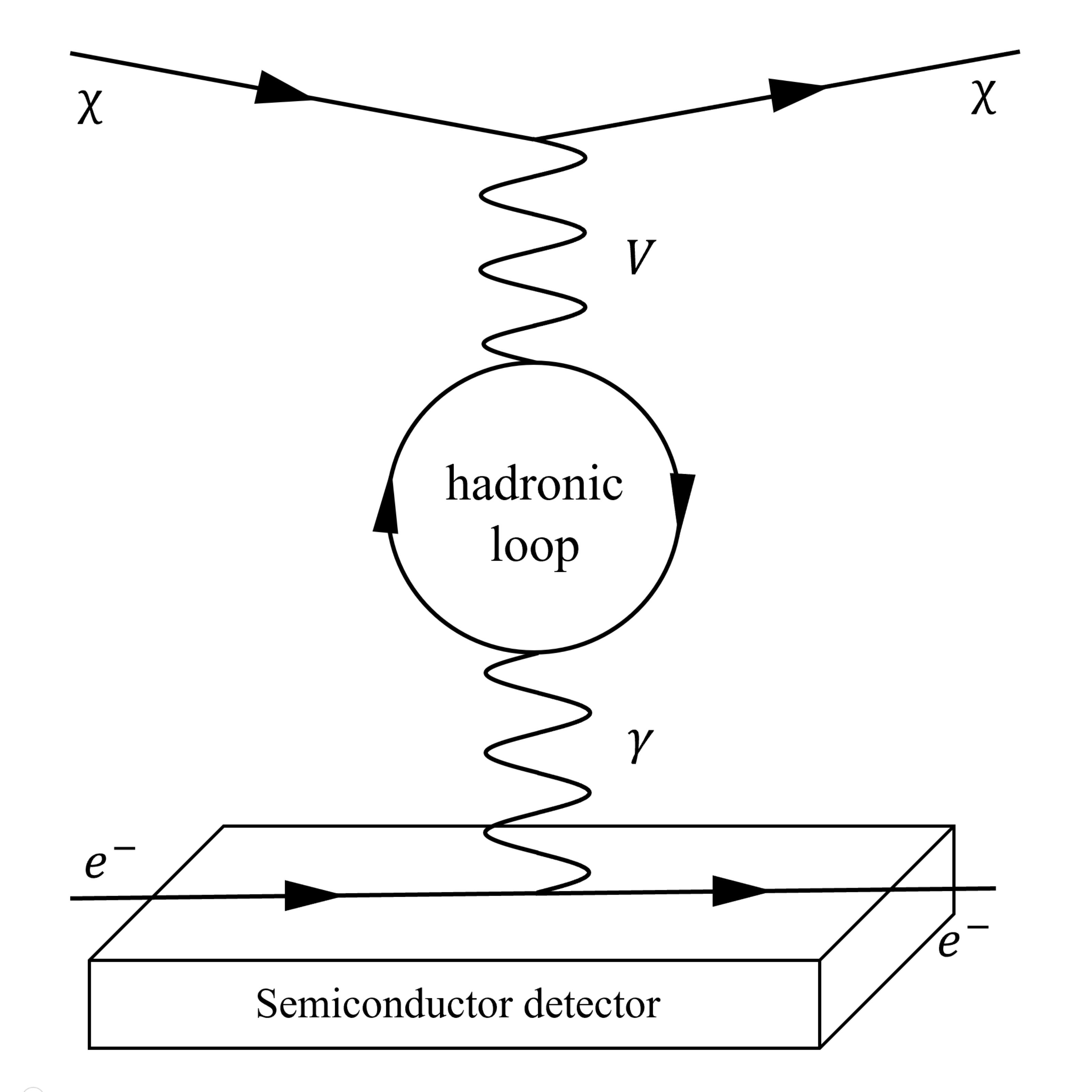}
    \caption{Schematic diagram of LDM-electron scattering process through a hadronic loop in a semiconductor detector.} 
    \label{fig:diagram}
\end{figure}

When LDM particles reach the semiconductor detectors, they can not directly couple to electrons at tree level as the mediator is leptophobic. However, their interaction can arise indirectly through hadronic loops, as shown in Fig.~\ref{fig:diagram}. The Dirac fermion LDM $\chi$ interact with standard model particles through a vector mediator $V$ that only couples to quarks. The simplest model is to introduce the vector mediator as the gauge boson associated with a local $U(1)_B$ baryon-number symmetry breaking at TeV or electroweak scale, as shown in various UV completions \cite{Nelson:1989fx,Rajpoot:1989jb,Foot:1989ts,He:1989mi,FileviezPerez:2011pt,Duerr:2013dza,FileviezPerez:2016laj}. Within an appropriate low-energy effective theory~\cite{Batell:2014yra}, the Lagrangian can be written as
\begin{equation}
    \mathcal{L} \supset g_B V_\mu J_B^\mu 
    + g_\chi V_\mu \bar{\chi} \gamma^\mu\chi,
\label{eq:Lagrangian}
\end{equation}
where $J_B^\mu = \frac{1}{3} \sum_q \bar{q}\gamma^\mu q$ is the corresponding baryon current. The parameters $g_\chi$ and $g_B$ denote the couplings of DM and quarks to the vector mediator, respectively. The relevant gauge anomalies can be canceled by extra fermions in the UV completion that acquire masses through the couplings to the spontaneous symmetry breaking sector of $U(1)_B$. These fermions are usually heavy and in agreement with experimental bounds \cite{Dobrescu:2014fca}. Depending on the gauge coupling, the gauge boson $V$ can be as light as the eV scale \cite{Fayet:1993fu,Fayet:2007ua,Nesvizhevsky:2007by,Bogorad:2023zmy}. In the energy regime of interest, the effective LDM-nucleon interaction Lagrangian is given by 
\begin{equation}
    \langle N | g_B  V_\mu J_B^\mu|N\rangle = g_B V_\mu \bar{N} \gamma^\mu N.
\end{equation}

The proton or meson loops can induce a vector mediator-photon mixing, which provides an effective coupling of LDM to electrons in semiconductor detector, as shown in Fig~\ref{fig:diagram}. In this framework, the scattering amplitude $\mathcal{M}_{i\to f}$ of a bound electron transition from the ground state $|i\rangle$ to the final state $|f\rangle$ with energy $\omega_f$ induced by LDM-electron scattering is written as
\begin{equation}
\begin{aligned}
\mathcal{M}_{i\to f} &=
g_{\chi}\bar{u}_{\chi}(p_{\chi }^\prime)\gamma^{\mu}u_{\chi}(p_{\chi})
\frac{g_{\mu\nu}}{m_{V}^2-q^2}
\frac{g_{\omega\xi}}{q^2}\\
&\times \Pi_{\nu\omega}^{\mathrm{loop}}\bar{u}_e(p_e^{\prime})\gamma^\xi u_e(p_e)\\
&=\bar{u}_{\chi}(p_{\chi }^\prime)\gamma^0 u_{\chi}(p_{\chi})
\frac{eg_{\chi}}{{m_{V}^2-q^2}}\left(\frac{c_{\mathrm{loop}}}{48\pi^2}\right)\\
&\times\langle f|\hat{\rho}(\mathbf{Q})|i\rangle2\pi\delta(\omega_f-\omega),\label{eq:amplitude}  
\end{aligned}
\end{equation}
where $m_V$ is the vector mediator mass. And the transfer four-momentum is represented as $q = (\omega, \mathbf{Q})$, where $\omega$ and $\mathbf{Q}$ are the transfer energy and three-momentum. The factor $\Pi_{\nu\omega}^{\mathrm{loop}}$ denotes the contribution of the hadronic loop, which is given by 
\begin{equation}
    \Pi_{\nu\omega}^{\mathrm{loop}}=  \frac{e}{48\pi^2} g_{\nu\omega}q^2c_{\mathrm{loop}},
\end{equation}
with
\begin{equation}
\begin{aligned}
c_{\mathrm{loop}} &= 4g_B\mathrm{ln}\left(\frac{4\pi e^{-\gamma_E}\Lambda^2}{m_p^2}\right),\label{eq:cloop}
\end{aligned}
\end{equation}
here $m_p = 938 \ \mathrm{MeV}$ is the proton mass, $e$ is the elementary charge. In this work, we assume universal mediator couplings to quarks, such that the dominant hadronic loop contribution is generated exclusively by the proton loop. Consequently, the contributions from the dominant light-meson loops, including pion and kaon loops, cancel identically (see Ref.~\cite{Diamond:2023fsm} for details). Following the standard treatment in baryon chiral effective field theory (see, e.g., Ref.~\cite{Becher:1999he}), we take the ultraviolet cutoff to be of the order of the hadronic (or chiral symmetry-breaking) scale, $\Lambda \sim \mathcal{O}(1~\mathrm{GeV})$, and choose $\Lambda = m_p$ as the benchmark value in our numerical calculations. We further set the Euler--Mascheroni constant to $\gamma_E = 0.577$.

In the second line of Eq.~\ref{eq:amplitude}, only the time-like component is retained, since spatial contributions are suppressed in the nonrelativistic bound state regime of the semiconductor electron, as discussed in Ref.~\cite{Liang:2024xcx}. The electron density operator in momentum space, $\hat{\rho}(\mathbf{Q})$, is related to the energy loss function (ELF) of semiconductor, 
\begin{equation}
    \mathrm{Im}\left(-\frac{1}{\epsilon\left(\mathbf{Q},\omega\right)}\right) =\frac{\pi e^{2}}{Q^{2}}\sum_{f}\left|\langle f|\hat{\rho}(\mathbf{Q})|0\rangle\right|^{2}\delta\left(\omega_{f}-\omega\right),
\end{equation}
where $\epsilon(\mathbf{Q}, \omega)$ is the material dielectric function, and is isotropic for the silicon detector, i.e., $\epsilon(\mathbf{Q}, \omega) = \epsilon({Q}, \omega)$. In this work, we adopt the ELF results of \textit{DarkELF} package~\cite{Knapen:2021bwg}.

With the scattering amplitude, the transition rate of electronic excitation induced by LDM in the silicon detector can be obtained as
\begin{equation}
    \begin{aligned}
        \Gamma\left({p}_{\chi}\right)& =
\sum_f\int\frac{\mathrm{d}^{3}p_{\chi}^{\prime}}
{\left(2\pi\right)^{3}}\frac{\left|\mathcal{M}_{i\to f}\right|^{2}}{4E_{\chi}E_{\chi}^{\prime}}
2\pi\delta\left(\omega+E_{\chi}^{\prime}-E_{\chi}\right)\\&=\int\frac{\mathrm{d}^3\mathbf{Q}}{\left(2\pi\right)^3}
\left(\frac{(2E_\chi-\omega)^2-Q^2}{4E_\chi(E_\chi-\omega)}\right)
2\frac{Q^2}{e^2}
\\& \times \mathrm{Im}\left(-\frac{1}{\epsilon\left({Q},\omega\right)}\right) \left[\frac{e g_\chi c_{loop}}{48\pi^2(Q^2-\omega^2+m_{V}^2)}\right]^2\\
&= \int\frac{\mathrm{d}^3\mathbf{Q}}{\left(2\pi\right)^3}
\frac{\pi \bar{\sigma}_{\chi n} \left [(2E_\chi-\omega)^2-Q^2\right]Q^2}{2\mu_{\chi n}^2 E_\chi(E_\chi-\omega)}\\
&\times \left( \frac{c_{\mathrm{loop}}}{48 \pi^2 g_B} \right)^2  \left|F_{\mathrm{DM}}(q)\right|^2
\mathrm{Im}\left(-\frac{1}{\epsilon\left({Q},\omega\right)}\right),
\end{aligned}
\end{equation}
where $E_\chi (E_\chi^\prime)$ represents the energy of initial (final) LDM, $\mu_{\chi n}$ is the reduced mass of DM and nucleon, and the DM form factor in the last line is defined as
\begin{equation}
    \left|F_{\mathrm{DM}}(q)\right|^{2} =\left\{\begin{array}{ll}
1 & \text { heavy med } \\
\left(\frac{\alpha m_e}{q}\right)^4 & \text { light med }
\end{array}\right.,
\end{equation}
with $m_e$ is the electron mass and $\alpha$ is the fine structure constant. In this work, we focus on the light mediator case, $m_V^2 \ll q^2$, where the DM form factor is enhanced at low momentum transfer in semiconductor detectors. Meanwhile, we define a momentum-independent LDM-nucleon cross section $\bar{\sigma}_{\chi n} \equiv \frac{\mu_{\chi n }^2  g_\chi^2 g_B^2}{\pi (\alpha^2m_e^2+m^2_{V})^2}$.

Combined with the differential flux of LDM $\mathrm{d}\Phi_\chi/\mathrm{d}T_\chi$, the event rate of electronic excitation induced by LDM in semiconductors can be expressed as
\begin{equation}
\begin{aligned}
    \frac{\mathrm{d}R}{\mathrm{d}\omega}
 &= \int \frac{\mathrm{d}T_\chi}{\rho_T}   \int \frac{\mathrm{d}\Omega}{4\pi} \frac{\mathrm{d}\Phi_\chi}{\mathrm{d}T_\chi} \frac{E_\chi}{|\vec{p}_\chi|}\Gamma\left(p_\chi\right)\delta\left(E_\chi'-E_\chi+\omega\right)\\
& =\frac{\bar{\sigma}_{\chi n} }{2\pi \rho_T} \frac{c_{\mathrm{loop}}^2}{ (48\pi^2\mu_{\chi n} g_B)^2 }  \int Q^3\mathrm{d}Q\int_{E_{\chi}^{\min}}\mathrm{d}E_\chi\frac{\mathrm{d}\Phi_\chi}{\mathrm{d}T_\chi} \\
    & \times \frac{(2E_\chi-\omega)^2-Q^2}{4v_\chi^2E_\chi^2}
    \left| F_{\mathrm{DM}}(q)\right|^2 \mathrm{Im}\left[\frac{-1}{\epsilon\left(Q,\omega\right)}\right], 
\label{eq:differential event rate}
\end{aligned}
\end{equation}
where $\rho_T$ is the mass density of target in solid detectors. The parameter $E_{\chi}^{\min} = \sqrt{(|\vec{p}_\chi| - |\vec{q}|)^2 +m_{\chi}^2} + \omega $ denotes the minimum LDM energy required to excite an electron signal with energy $\omega$. 

However, direct plasmon excitation requires additional kinematic conditions. For instance, in the case of DM-electron scattering via a dark photon mediator, the DM velocity must exceed $0.01c$~\cite{Liang:2024xcx}. For LDM-electron scattering via a hadronic loop, the vector mediator--photon mixing induced by the loop does not alter the four-momentum transfer before and after the loop, making the situation analogous to the dark photon case. Consequently, the kinematic condition for direct plasmon excitation induced by LDM is the same as in Ref.~\cite{Liang:2024xcx}, i.e., $v_\chi \gtrsim 0.01c$. In this work, we therefore consider two mechanisms capable of producing relativistic DM, boosted DM in two-component DM model and cosmic-ray up-scattering DM (CRDM).

\textit{ \textbf{1) Boosted DM in two-component DM model}:} A commonly studied framework for relativistic DM is boosted DM in the two-component DM model. In this model, the DM consists of two species, $\chi_1$ and $\chi_2$, with the masses $m_{\chi_1} > m_{\chi_2}$. The lighter DM $\chi_2$ can be as light as the eV scale and its population may contribute to the energy density of radiation in the early universe and alter the effective number of neutrinos during Big
Bang Nucleosynthesis (BBN). However, the two-component DMs can decouple around the QCD phase transition to suppress the temperature $T_D$ of the dark sector, relative to the neutrino
temperature $T_\nu$, due to the annihilation of degrees of freedom in the Standard Model (SM) plasma \cite{Cherry:2015oca,McKeen:2018pbb}. Therefore, the contribution of the relativistic component $\chi_2$ to the change of the effective number of neutrinos $\Delta N_{\rm eff}$ during BBN is suppressed by $({T_D}/{T_\nu})^4$ and compatible with current constraints \cite{Cherry:2015oca}. The heavier DM $\chi_1$, which constitutes the dominant DM component, does not couple directly to SM particles and achieves the correct relic abundance through annihilation into the lighter DM, $\chi_1 \bar{\chi}_1 \rightarrow \chi_2 \bar{\chi}_2$. The mass hierarchy between the two species imparts a Lorentz boost to $\chi_2$, with $\gamma_m = m_{\chi_1}/m_{\chi_2}$. Thus, the lighter DM $\chi_2$ is often referred to as boosted dark matter (BDM), which can be relativistic depending on the mass splitting $\Delta m_{\chi} = m_{\chi_1} - m_{\chi_2} $. In this work, the BDM $\chi_2$ interact with SM particle via the Lagrangian in Eq.~\ref{eq:Lagrangian}.

The differential flux of BDM $\chi_2$ from the galactic center (GC) can be calculated  by
\begin{equation}
    \frac{\mathrm{d}\Phi_{\mathrm{BDM}}}{\mathrm{d}E_{\chi_{2}}} = 
    \frac{1}{4\pi} \frac{\langle \sigma v \rangle}{4 m_{\chi
    _{1}}^2}
    J(\Omega) 2 \delta(E_{\chi_2} - m_{\chi_1})
\end{equation}
where the $\langle \sigma v \rangle$ denotes the thermally-averaged annihilation cross section of $\chi_1 \bar{\chi}_1 \rightarrow \chi_2 \bar{\chi}_2$, and the term $2\,\delta(E_{\chi_2} - m_{\chi_1})$ indicates that the annihilation process produces two monoenergetic BDM particles $\chi_2$. The astrophysical $J(\Omega)$ factor is given by the three-dimensional integral of DM density over the target solid angle in the sky,
\begin{equation}
    J(\Omega) = \int \mathrm{d} \Omega \int_{\mathrm{l.o.s}} \rho_\chi^2(r (x,\theta))\mathrm{d}x,
\end{equation}
where $\rho_\chi(r)$ denotes the DM density profile, where the Galactocentric distance is given by 
$r = \sqrt{d_{\odot}^2 + x^2 - 2x d_{\odot} \cos \theta}$.
$x$ is the line-of-sight (l.o.s.) distance from the Earth, $d_{\odot} = 8.127~\mathrm{kpc}$ is the distance from the Sun to the GC, and $\theta$ is the angle between the Earth-GC axis and the l.o.s.

In this work, we adopt the Navarro-Frenk-White (NFW) DM density profile and the results in Ref.~\cite{Agashe:2014yua}, the flux of BDM form the all-sky is given by
\begin{equation}
\begin{aligned}
\frac{\mathrm{d} \Phi_{\mathrm{BDM}}}{\mathrm{d}E_{\chi_{2}}}&=4.0\times10^{-7}\mathrm{cm}^{-2}\mathrm{s}^{-1}
\left( \frac{\langle \sigma v \rangle}{5 \times10^{-26} \;\mathrm{cm}^3\mathrm{s}^{-1}}  \right)\\
&\times \left(\frac{20\;\mathrm{GeV}}{m_{\chi_1}}\right)^2 \delta(E_{\chi_2} - m_{\chi_1}).
\label{eq:BDMflux}     
\end{aligned}
\end{equation}

\textit{ \textbf{2) Cosmic-ray up-scattering DM}:} 
Apart from the BDM in two-component DM model, another popular relativistic component of the DM flux in recent years arises from cosmic-ray up-scattering DM, which is short for CRDM. In this framework, the DM halo in the Milky Way can be accelerated by the scattering between high-energy cosmic-ray and DM.

Therefore, the differential flux of CRDM can be written as
\begin{equation}
\begin{aligned}
     \frac{\mathrm{d}\Phi_{\mathrm{CRDM}}}{\mathrm{d}T_\chi} & = \int\frac{\mathrm{d}\Omega}{4\pi} \int_{\mathrm{l.o.s}}^{\infty} \mathrm{d}s \int _{T_p^{min}}^{\infty}\mathrm{d}T_p \frac{\rho_\chi}{m_\chi}\frac{\mathrm{d}\sigma_{\chi n}}{\mathrm{d}T_\chi}\frac{\mathrm{d}\Phi_p}{\mathrm{d}T_p}    \\
     &=D\frac{\rho_0}{m_\chi}\int_{T_p^{\mathrm{min}}}^\infty\mathrm{d}T_p\frac{\mathrm{d}\sigma_{\chi n}}{\mathrm{d}T_\chi}\frac{\mathrm{d}\Phi_p^{\mathrm{LIS}}}{\mathrm{d}T_p},
\end{aligned} 
\end{equation}
with the differential cross section of DM-proton elastic scattering in the rest frame of the DM particles,
\begin{equation}
    \begin{aligned}
    \frac{\mathrm{d}\sigma_{\chi n}}{\mathrm{d}T_\chi} 
    &= \frac{\bar{\sigma}_{\chi n}|F_{\mathrm{DM}}(q)|^2F_p^2(q^2)}{4\mu_{\chi p}^2(2m_p T_p + T_p^2)}[ 2m_\chi(T_p + m_p)^2  \\ &+m_\chi T_\chi^2- 2m_\chi T_p T_\chi -T_\chi(m_p + m_\chi)^2 ],
    \end{aligned}
\end{equation}
where $T_p$ and $T_\chi$ are the kinetic energy of DM  and CR proton, respectively.  For light nuclei such as the proton, the nucleon form factor is often parameterized by the dipole form 
\begin{equation}
    F_p(q^2) = \frac{1}{(1+q^2/\Lambda_p^2)^2},
\end{equation}
where $\Lambda_p \simeq 770~\mathrm{MeV}$ is related to the proton charge radius. The minimal kinetic energy of the proton is 
\begin{equation}
T_{p}^{\min }= \left(\frac{T_{\chi}}{2}-m_{p}\right)\left[1\pm \sqrt{1+\frac{2 T_{\chi}}{m_{\chi}} \frac{\left(m_{p}+m_{\chi}\right)^{2}}{\left(2 m_{p}-T_{\chi}\right)^{2}}}\right],
\end{equation}
where the $+$ ($-$) sign denotes the $T_{\chi} > 2m_p$ ($T_{\chi} < 2m_p$). For the $T_{\chi} = 2m_p$, the minimal kinetic energy is given by $\sqrt{{m_{p}}/{m_{\chi}}}\left(m_{p}+m_{\chi}\right)$.

In this work, we consider only the contribution from the local interstellar (LIS) proton spectrum, which can be parameterized as $\mathrm{d}\Phi_p^{\mathrm{LIS}}/\mathrm{d}T_p = 4\pi \, (\mathrm{d}R_p/\mathrm{d}T_p)(\mathrm{d}I_p/\mathrm{d}R_p)$, with $R_p$ denoting the CR proton rigidity~\cite{Boschini:2017fxq}.  The contribution from DM--electron scattering is suppressed by the proton loop and can therefore be neglected. Meanwhile, we assume a homogeneous CR distribution and define the effective distance as $D \equiv \frac{1}{4\pi \rho_0} \int_{\mathrm{l.o.s}} \rho_\chi \,\mathrm{d}s \,\mathrm{d}\Omega $, which evaluates to $D = 8.02~\mathrm{kpc}$ under the assumption of an NFW DM profile with local density $\rho_0 = 0.4~\mathrm{GeV}/\mathrm{cm}^3$ and a l.o.s integration up to 10 kpc~\cite{Bringmann:2018cvk}. An inhomogeneous and larger-volume CR distribution would result in a larger effective distance 
and thus a higher CRDM flux~\cite{Cappiello:2019qsw, Xia:2022tid}, however, which is beyond the scope of this work.



\section{Result}

\begin{figure}[htbp]
  \centering
\includegraphics[height=6cm,width=8cm]{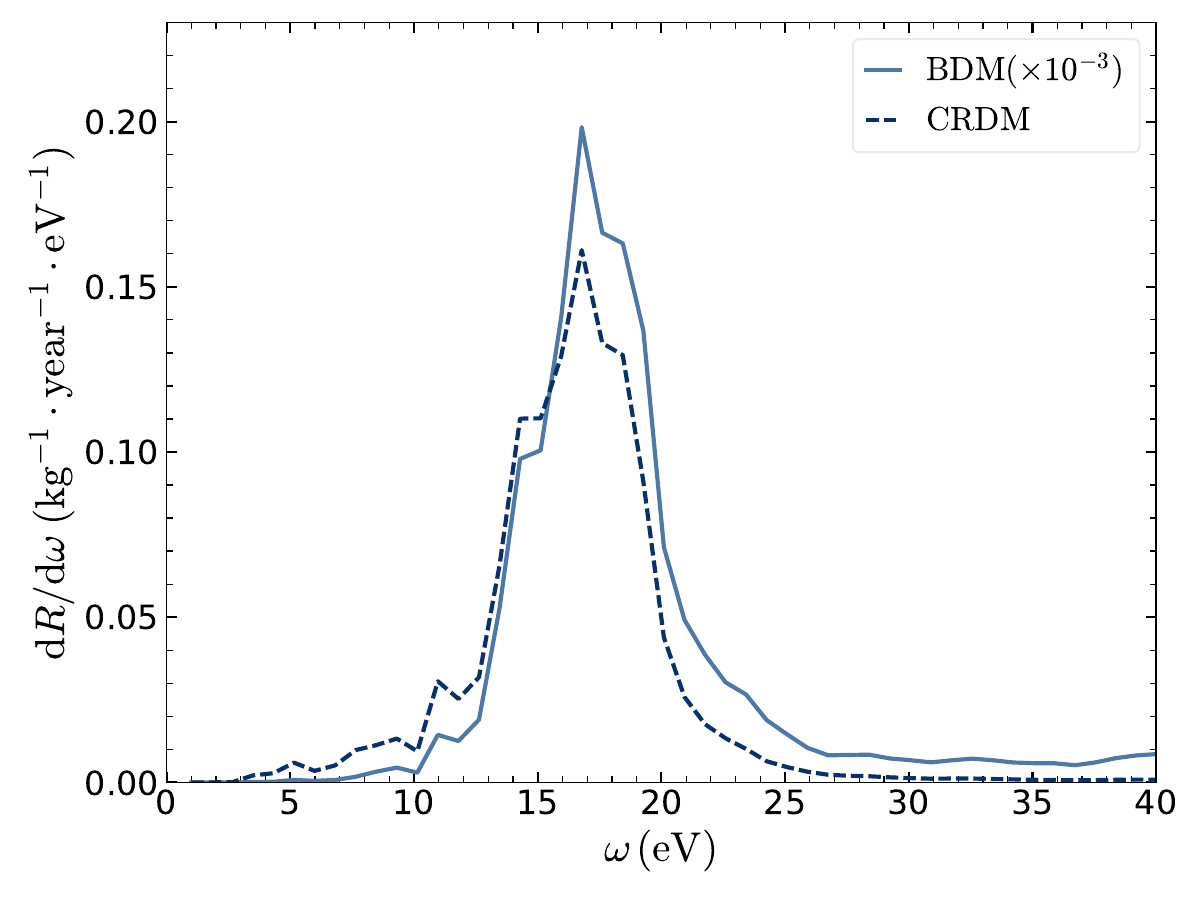}
\includegraphics[height=6cm,width=8cm]{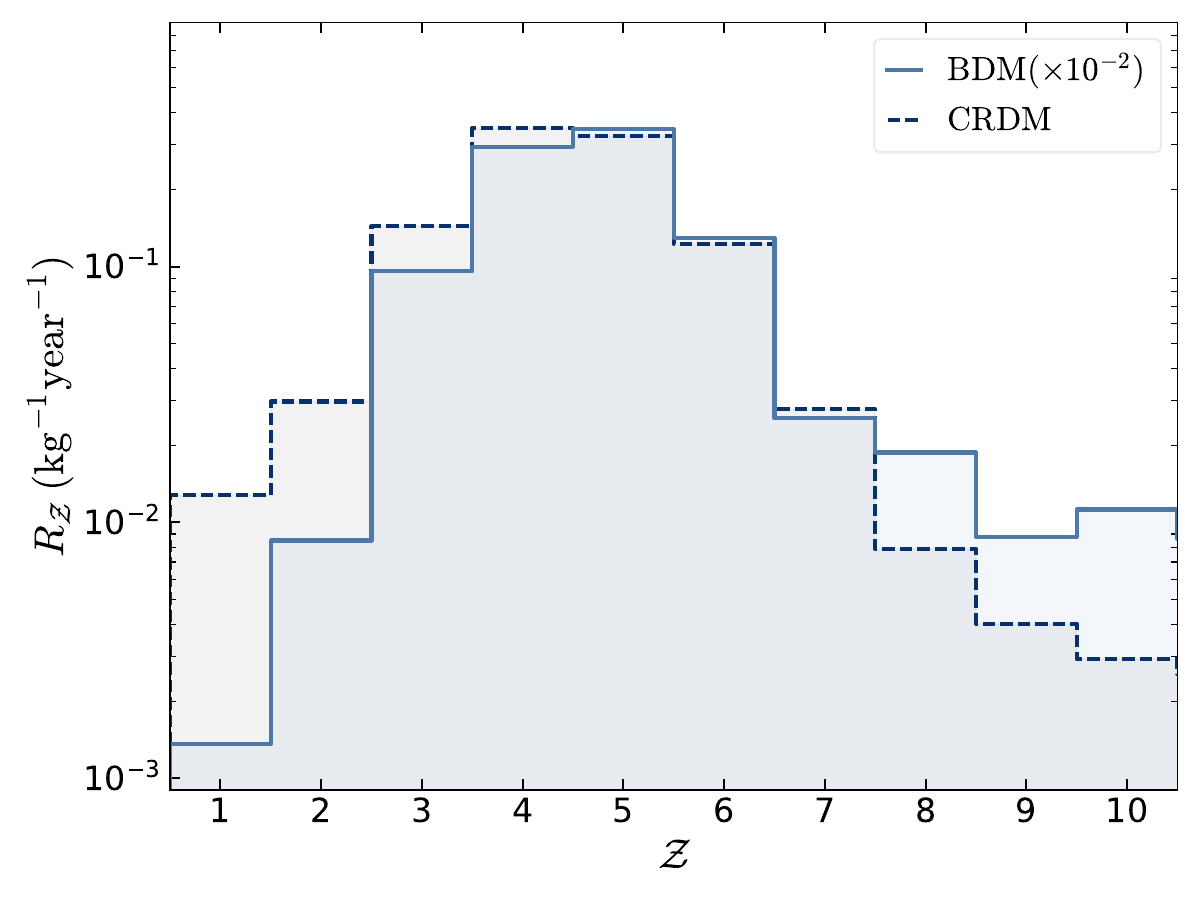}
\caption{ {\it Top panel:} The differential event rate $\mathrm{d}R/\mathrm{d}\omega$ as a function of the deposited energy $\omega$; {\it Bottom panel:} Event rate $R_{\mathcal{Z}}$ as a function of the ionized charge $\mathcal{Z}$. The solid lines denotes the BDM with $m_{\chi_{1}}=1$ MeV, $m_{\chi_{2}} = 1$ keV and thermally averaged the annihilation cross section $\langle \sigma v \rangle = 5\times 10^{-26} \mathrm{cm^3/s}$. The dashed lines are the results of 1 keV CRDM. The brenchmark DM-nucleon scattering cross section is  $\bar{\sigma}_{\chi n} = 10^{-32} \; \mathrm{cm}^2$.}
\label{fig:dRdomega}
\end{figure}

To more clearly illustrate the electronic excitation induced by BDM and CRDM in a silicon detector, the top panel of Fig.~\ref{fig:dRdomega} shows the differential event rate $\mathrm{d}R/\mathrm{d}\omega$ for 1 keV BDM (solid line: $m_{\chi_{1}} = 1\; \mathrm{MeV}$, $m_{\chi_{2}} = 1\; \mathrm{keV}$) and CRDM (dashed line), assuming a DM--nucleon scattering cross section of $\bar{\sigma}_{\chi n} = 10^{-32}\; \mathrm{cm}^2$. It is evident that the event rates exhibit a pronounced enhancement in the region 
$10\ \mathrm{eV} < \omega < 25\ \mathrm{eV}$ for both BDM and CRDM, consistent with the plasmon excitation 
region shown in Fig.1 of Ref.~\cite{Liang:2024xcx} and with observations from electron energy loss spectroscopy~\cite{1988PhDT15K}.

For the semiconductor detector such as SENSEI and DAMIC-M experiment, the electronic excitation signal is presented in the secondary electron-hole pairs observation,  and its relationship with the differential event rate is as follows
\begin{equation}
    R_{\mathcal{Z}} = \int \mathrm{d}\omega P(\mathcal{Z}|\omega)\frac{\mathrm{d}R}{\mathrm{d}\omega},
\label{eq:event_n}
\end{equation}
where $\mathcal{Z}$ denotes the ionized charge of secondary electron-hole pairs. The function $P(\mathcal{Z}|\omega)$ denotes the pair-creation probability distribution, which represents the probability of producing $\mathcal{Z}$ ionization pairs for a given deposited energy $\omega$. Moreover, this probability is temperature-dependent. In this work, we use the $P(\mathcal{Z}|\omega)$ results at $100~\mathrm{K}$ from Ref.~\cite{Ramanathan:2020fwm}, which is close to the operating temperatures of the SENSEI and DAMIC-M experiments. The corresponding band gap energy and mean energy per pair are $E_{\mathrm{gap}} = 1.16$ eV and $\epsilon = 3.75$ eV, respectively. Thus, for the study of plasmon excitation, we focus on the deposited energy $\omega$ range from $E_{\mathrm{gap}}$ up to $50~\mathrm{eV}$. The bottom panel of Fig.~\ref{fig:dRdomega} shows the event rate $R_{\mathcal{Z}}$ 
as a function of the ionized charge $\mathcal{Z}$. The plasmon excitations induced by BDM and CRDM via a hadronic loop are found to lie in the range of 3 to 6 electron-hole pairs. 

The SENSEI experiment has recently reported multi-electron results, observing four $3e^{-}$ events and zero events in the $4$-$6e^{-}$ range, with an exposure of unmasked pixels is $100.72$ gram$\cdot$days~\cite{SENSEI:2023zdf}. The corresponding 90\% confidence level (C.L.) upper limits on the number of events and the effective exposures are summarized in Table~\ref{tab:experiment data}. Furthermore, the DAMIC-M experiment has reported one $4e^{-}$ event with an exposure of $1.3~\mathrm{kg}\cdot\mathrm{days}$~\cite{DAMIC-M:2025luv}, which is consistent with the results of the SENSEI experiment. However, we do not present the corresponding constraints from DAMIC-M experiment in this work due to the lack of effective exposure information.

\begin{table}[htbp]
    \centering
    \begin{tabular}{c|c|c}
    \hline
    Ionized Charge $\mathcal{Z}$  & Exposures (gram$\cdot$days) & Event Number \\
    \hline
    $ 3e^-$ & 57.71 & 8.6\\
    $ 4e^-$ & 63.03 & 2.3\\
    $ 5e^-$ & 65.56  & 2.3\\
    $ 6e^-$ & 67.31  & 2.3\\
    \hline
    \end{tabular}
    \caption{The 90\% C.L. upper limits on the number of observed events in the $3$-$6e^{-}$ range from the SENSEI experiment~\cite{SENSEI:2023zdf}.}
    \label{tab:experiment data}
\end{table}

\begin{figure}[ht]
  \centering
\includegraphics[height=7.2cm,width=8cm]{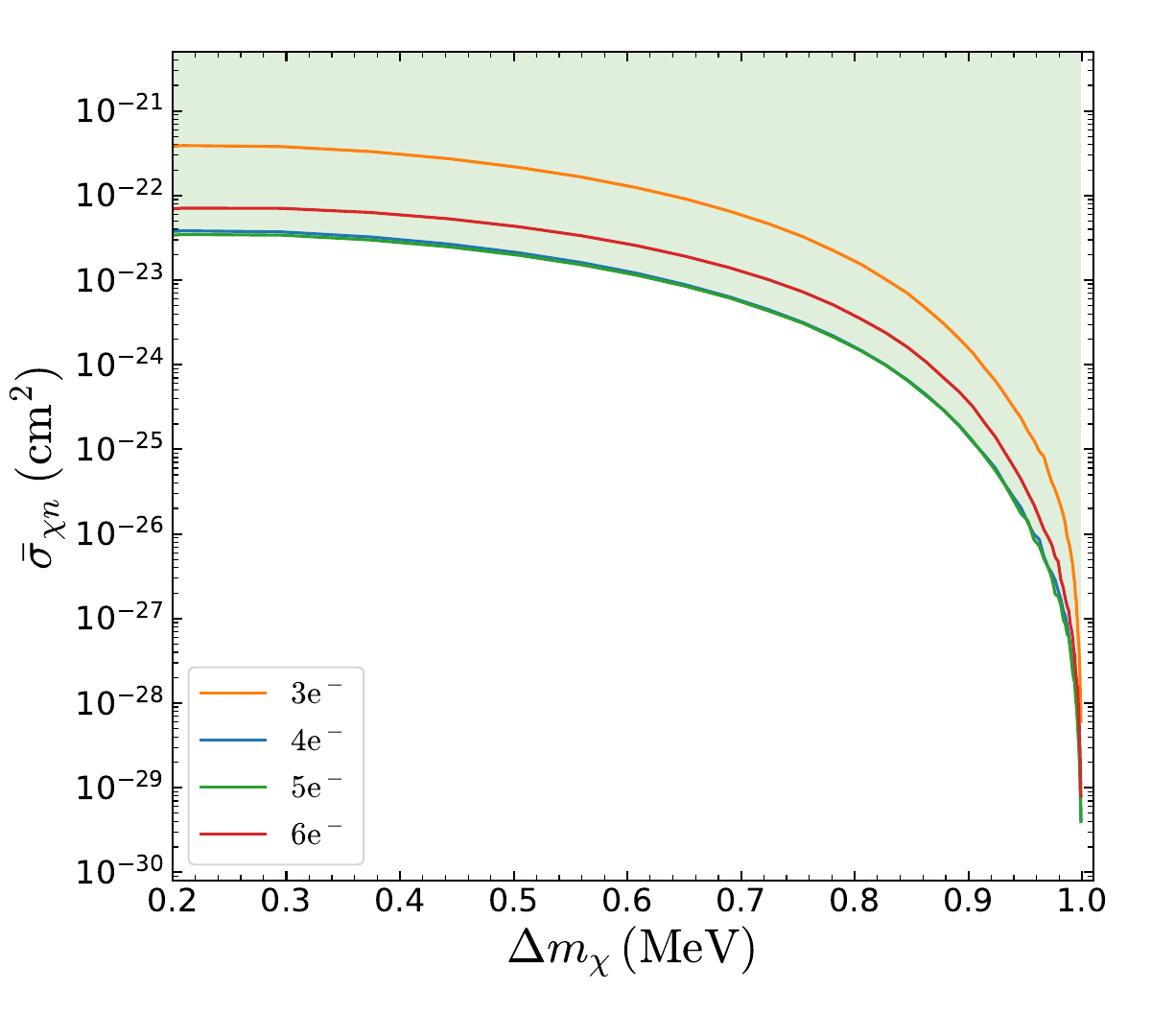}
\hspace{0.02\textwidth}
\includegraphics[height=7.2cm,width=8cm]{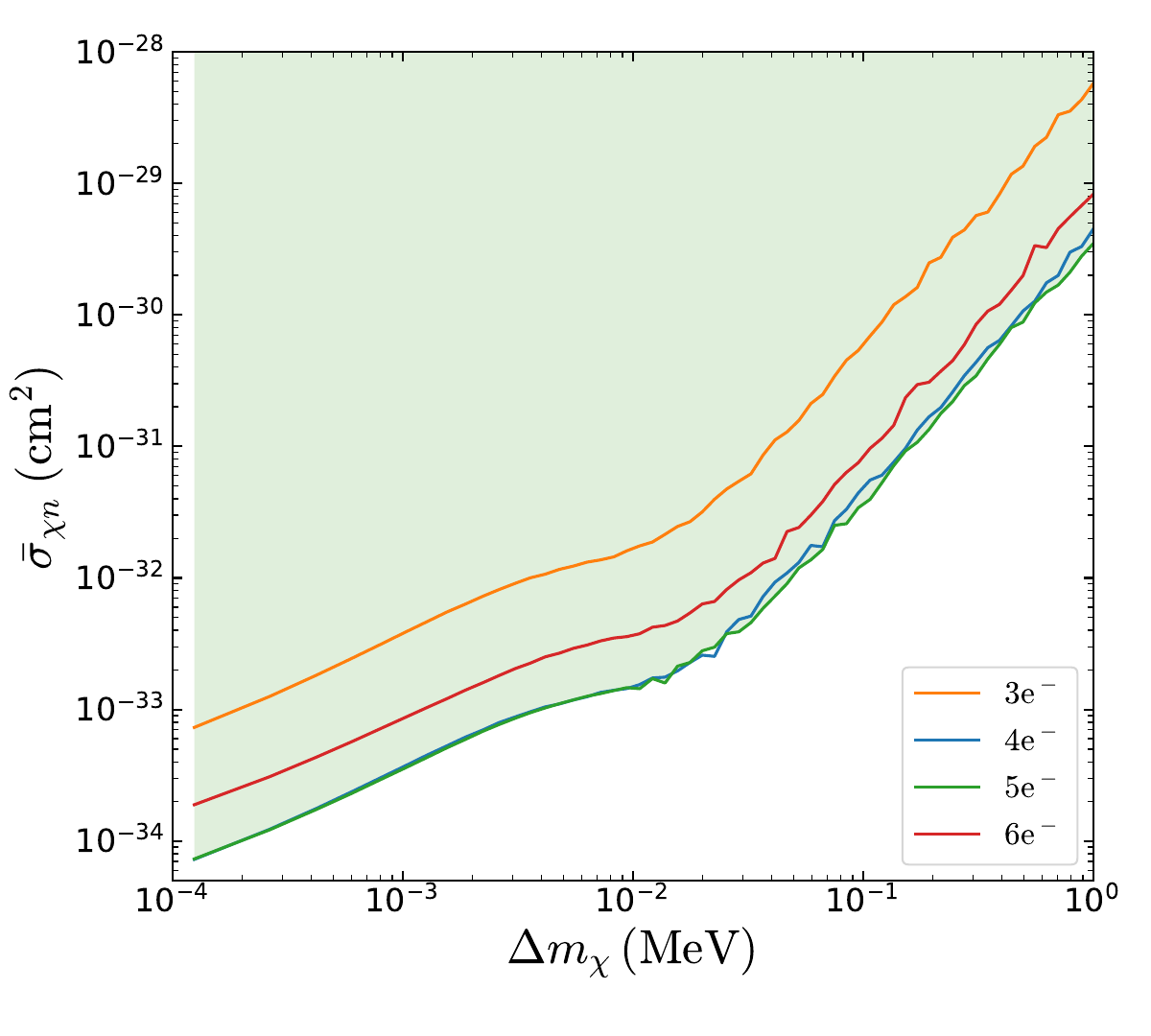}
\caption{The 90\% confidence level exclusion limits on the LDM-nucleon scattering cross section $\bar{\sigma}_{\chi n}$ as a function of the mass splitting $\Delta m_{\chi} = m_{\chi_1} - m_{\chi_2}$ in the two-component BDM model. {\it Top panel:} results with the heavier component mass fixed at $m_{\chi_1} = 1~\mathrm{MeV}$. {\it Bottom panel:} results with the lighter component mass fixed at $m_{\chi_2} = 1~\mathrm{keV}$.}
\label{fig:mass splitting}
\end{figure}

By combining the experimental results with the theoretical event rate in Eq.~\ref{eq:event_n}, we derive the 90\% C.L. exclusion limits by requiring the predicted number of events not to exceed the observed limits. For the BDM scenario in the two-component model, we first present the exclusion limits from each 3-6$e^{-}$ bin in the $\bar{\sigma}_{\chi n}$-$\Delta m_{\chi}$ plane, respectively, as shown in Fig.~\ref{fig:mass splitting}, to investigate the impact of the boost factor of BDM $\chi_2$. In the top panel, the mass of the heavier component is fixed at $m_{\chi_1} = 1~\mathrm{MeV}$ while the lighter component mass $m_{\chi_2}$ is varied over the range from the keV to MeV scale. Here we find that the constraints become weaker as $\Delta m_{\chi}$ decreases (corresponding to a smaller $m_{\chi_2}$). This behavior arises because the electronic excitation rate in Eq.~\ref{eq:differential event rate} scales approximately with the inverse square of the reduced mass, $R \propto 1/\mu_{\chi n}^2 \sim 1/m_{\chi_2}^2$. Conversely, in the bottom panel the lighter component is fixed at $m_{\chi_2} = 1~\mathrm{keV}$ and the mass of the heavier component is varied. The mass of the heavier component affects the electronic excitation rate not only through the BDM flux (Eq.~\ref{eq:BDMflux}), but also via the factor $S=\tfrac{(2E_{\chi}-\omega)^2-Q^2}{4v_{\chi}^2 E_\chi^2}$ in Eq.~\ref{eq:differential event rate}. For the case $\Delta m_{\chi} \gg m_{\chi_2}$, the factor $S$ remains approximately constant. However, as $\Delta m_{\chi}$ or $m_{\chi_1}$ decreases, $S$ factor becomes suppressed by $m_{\chi_1}$, leading to a change in the slope of the exclusion limits around $\Delta m_{\chi} \sim 20~\mathrm{keV}$. This trend ceases once the mass splitting becomes too small to produce (semi-)relativistic BDM, i.e., $v_{\chi} \gtrsim 10^{-2}c$. Besides, the results in Fig.~\ref{fig:mass splitting} indicate that the strongest constraint arises from the $5e^{-}$ observation in the SENSEI experiment. Therefore, in the following discussion we focus on the constraint from the $5e^{-}$ bin.

\begin{figure}[ht]
    \centering
    \includegraphics[height=7.2cm,width=8cm]{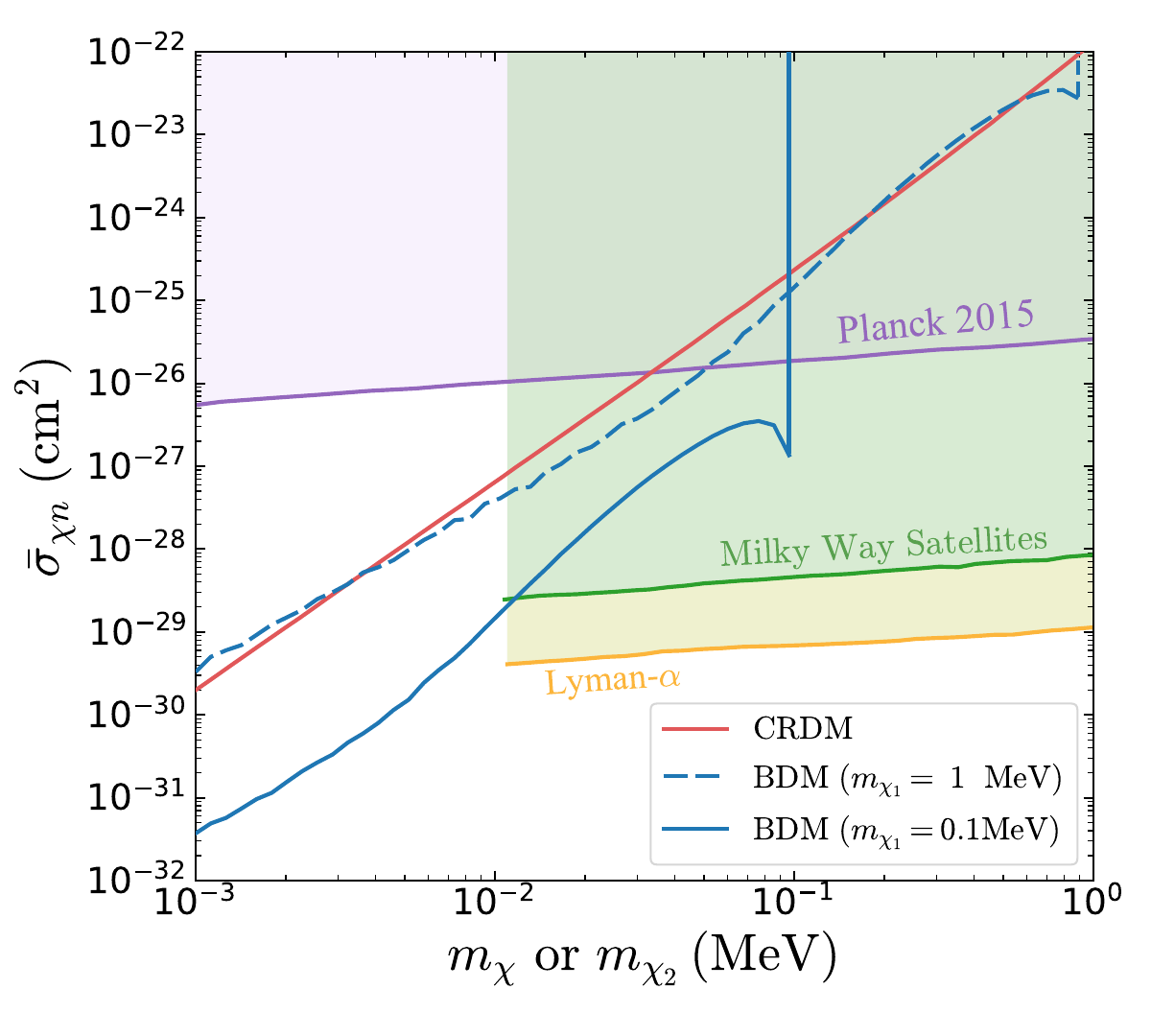}
    \caption{The 90\% C.L. upper limits on the LDM-nucleon scattering cross section $\bar{\sigma}_{\chi n}$ as a function of DM mass. The red line corresponds to CRDM, while the dashed and solid blue lines represent the two-component BDM scenario with $m_{\chi_1} = 1~\mathrm{MeV}$ and $m_{\chi_1} = 0.1~\mathrm{MeV}$, respectively. For comparison, existing cosmological and astrophysical constraints are also shown: CMB from \textit{Planck} 2015~\cite{Planck:2015bpv,Gluscevic:2017ywp} (purple), Milky Way satellite galaxies~\cite{Nadler:2019zrb} (green), and the Lyman-$\alpha$ forest~\cite{Rogers:2021byl} (yellow).     } 
    \label{BDMlimit}
\end{figure}

In Fig.~\ref{BDMlimit}, we present the 90\% C.L. constraints on the LDM-nucleon cross section $\bar{\sigma}_{\chi n}$ as a function of the DM mass $m_{\chi}$ (or $m_{\chi_2}$) for both the BDM and CRDM scenarios. For the two-component BDM case, we consider two benchmark values for the heavier component, $m_{\chi_1} = 1~\mathrm{MeV}$ (dashed blue line) and $m_{\chi_1} = 0.1~\mathrm{MeV}$ (solid blue line), motivated by the results in Fig.~\ref{fig:mass splitting}. As shown in the figure, smaller values of $m_{\chi_1}$ yield more stringent constraints on the BDM-nucleon scattering cross section, consistent with the trend observed in the bottom panel of Fig.~\ref{fig:mass splitting}. The exclusion limit from plasmon excitation induced by CRDM (red line) is similar to that obtained for the two-component BDM scenario with $m_{\chi_1} = 1~\mathrm{MeV}$. In addition, several cosmological and astrophysical constraints are shown in this parameter space, including the CMB from Planck 2015~\cite{Gluscevic:2017ywp,Planck:2015bpv} (purple), Milky Way satellite galaxies~\cite{Nadler:2019zrb} (green), and the Lyman-$\alpha$ forest~\cite{Rogers:2021byl} (yellow). It should also be noted that a large DM-nucleon cross section suppresses the DM flux reaching the detector due to interactions with the Earth, which will reduce the experimental sensitivity. A typical reference value for this cross section is $\mathcal{O}(10^{-28}) \; \mathrm{cm}^2$~\cite{Xia:2021vbz,Kumar:2024xfb}. Our results provide new exclusion limits on the LDM-nucleon scattering cross section based on the observation of electronic collective excitation. Finally, it is worth noting that the light mediator $V$ is subject to constraints from fifth-force searches~\cite{Murata:2014nra}, neutron scattering and neutron-optics experiments~\cite{Leeb:1992qf}, and other probes~\cite{Rrapaj:2015wgs,Berryman:2022zic}. For the light-mediator mass range $m_V\ll 1~\mathrm{keV}$ considered here, laboratory experiments constrain the gauge coupling to $g_B<10^{-9}$ at $m_V\sim20~\mathrm{eV}$ \cite{Berryman:2022zic}. In our work, for $m_{\chi_2}=1~\mathrm{keV}$ and $g_\chi=1$, the BDM limits on $g_B$ are $g_B<2.5\times10^{-6}$ and $2.5\times10^{-7}$ for
$m_{\chi_1}=1$ and $0.1~\mathrm{MeV}$, respectively. For CRDM with $m_\chi=1~\mathrm{keV}$, the corresponding limit is
$g_B<1.8\times10^{-6}$.

\section{Conclusions } 
In this work, we have investigated a kind of leptophobic dark matter models in which tree-level DM-electron interactions are absent, while hadronic loop effects can effectively induce DM-electron scattering channels. Such loop-induced interactions allow leptophobic dark matter to produce electronic excitation signals in dark matter detectors, in particular collective excitations (plasmons) in semiconductors, which in turn provide constraints on the DM-nucleon scattering cross section through electronic signal measurements. To demonstrate the potential reach of this mechanism, we considered two representative relativistic dark matter scenarios, boosted dark matter in a two-component model and cosmic-ray up-scattering dark matter. Both models naturally yield dark matter with velocities high enough to excite plasmons in semiconductor targets. Using currently available experimental data, we derived new 90\% confidence level exclusion limits on the spin-independent DM-nucleon cross section for dark matter masses ranging from the keV to the MeV scale. Our study shows that electronic collective excitations can serve as a powerful probe of leptophobic dark matter, extending the sensitivity of direct detection beyond the capabilities of conventional nuclear recoil based searches. Future detectors with lower thresholds and larger exposures will further enhance this potential, opening new opportunities to explore sub-GeV leptophobic dark matter through electronic signals.

\section*{Acknowledgments}

N. Liu is supported by the National Natural Science Foundation of China (NNSFC) under grants No. 12275134 and No. 12335005. L. Su is supported by the Alexander von Humboldt Foundation. B. Zhu is supported by the National Natural Science Foundation of China (NNSFC) under grant
No. 12275232.

\bibliography{reference}

\end{document}